\documentclass[aps,twocolumn,showpacs,prb]{revtex4}
\usepackage{dcolumn}
\usepackage{epsf}
\usepackage{amsmath}

\begin{document}

\title {Magnetotransport near a quantum critical point in a simple metal }
\author{Ya. B. Bazaliy$^a$}
\author{R. Ramazashvili$^a$}
\author{Q. Si$^b$}
\author{M. R. Norman$^a$}
\affiliation{ $^a$ Materials Science Division,
Argonne National Laboratory, Argonne, IL 60439; \\
$^b$ Department of Physics and Astronomy, \\
Rice University, Houston, TX 77005-1892}
\date{\today}

\begin{abstract}
We use geometric considerations to study transport properties, such as the
conductivity and Hall coefficient, near the onset of a nesting-driven spin density
wave in a simple metal. In particular, motivated by recent experiments on
vanadium-doped chromium, we study the variation of transport coefficients with the
onset of magnetism within a mean-field treatment of a model that contains nearly
nested electron and hole Fermi surfaces. We show that most transport coefficients
display a leading dependence that is linear in the energy gap. The coefficient of
the linear term, though, can be small. In particular, we find that the Hall
conductivity $\sigma_{xy}$ is essentially unchanged, due to electron-hole
compensation, as the system goes through the quantum critical point.  This
conclusion extends a similar observation we made earlier for the case of completely
flat Fermi surfaces to the immediate vicinity of the quantum critical point where
nesting is present but not perfect.

\end{abstract}
 \maketitle

\section{Introduction}

Continuous zero temperature phase transitions
have attracted much interest lately, due to a
variety of unusual low-temperature properties
observed in their vicinity. Experiments on a
broad range of materials, including cuprates,
heavy fermion compounds and transition metals,
have shown thermodynamic, magnetic and transport
properties that often do not fit in the framework
of the Landau Fermi liquid theory of metals. This
raises the possibility that the critical ground state
may be a new, exotic state of matter. In many cases,
understanding such a transition is rendered difficult
by the fact that the phases involved are not well
understood. This is the case for the cuprate
superconductors, where even the underlying `normal'
state remains a subject of some debate, and for many
heavy fermion materials,
which involve a nontrivial interplay between
the f-electron magnetic moments and conduction electron sea.
Thus, it is important
to study a case of a simple magnetic metal, where both
the magnetic and the paramagnetic phases, as well
as the nature of the transition,
are believed to be well understood.

The Cr$_{1-x}$V$_x$ alloy is an ideal candidate for such a study (see reviews
\cite{fawcett1,fawcett2}). Pure chromium is a textbook example of a nesting-driven
antiferromagnet, where doping with vanadium suppresses
the N\'eel temperature T$_N$
to zero already at $x = x_c \approx 3.5\%$. In recent experiments on
Cr$_{1-x}$V$_x$, \cite{rosenbaum2} T$_N$ has been driven to zero by pressure, which
allows access to the quantum critical point with minimal involvement of
doping-induced disorder.

Of various thermodynamic and transport properties near a quantum critical point, the
Hall transport attracted particular interest lately, as it carries information about
the change of the Fermi surface volume in an interacting system
\cite{revaz_science,lqcp}. Here we will stay within an SDW description, for which
the corresponding quantum critical point is above the upper critical dimension
\cite{hertz,millis}.
We will focus exclusively on the static transport coefficients and, with the
goal of treating the problem within the simplest possible model, confine
ourselves to a mean-field approach.
While the behavior of the electric conductivity near the onset of a density wave
within such an approach has been studied early on \cite{elliott+wedgwood}, the Hall
conductivity has not been investigated until recently, when it was discussed for an
SDW \cite{our_chromium_PRL} and for d-density wave order \cite{dwaveorder}. Standard
expressions \cite{ziman} give electric and Hall conductivities in terms of integrals
over the Fermi surface. When evaluating them, one has to take into account the Fermi
surface change in the ordered state, as well as the change of Fermi velocities and
mass tensors. Consequently, developing intuition about the behavior of these
quantities is non-trivial.

In our earlier work \cite{our_chromium_PRL}, we considered a model in which the
Fermi surface contains completely flat portions. In that case, the Hall number has a
discontinuity at the quantum critical point, but the Hall conductivity does not
change due to the zero curvature of the flat parts.  Moreover, if the tuning
parameter (such as pressure) does not change the elastic scattering, the Hall
coefficient should scale as the square of the longitudinal resistivity.  This
behavior was confirmed by detailed numerical simulations based on the actual Fermi
surface of chromium, and the results were in good agreement with Hall data
\cite{rosenbaum1} across the doping induced quantum phase transition in
Cr$_{1-x}$V$_x$.

Fermi surfaces, however, are never perfectly flat. The finite curvature becomes
especially important in the immediate vicinity of the quantum critical point, which
has become experimentally accessible with pressure in Cr$_{1-x}$V$_x$.
\cite{rosenbaum2} In this paper, we address this regime in some detail by
considering the Rice model \cite{rice}. The model, motivated by the electronic
structure of chromium, incorporates an electron Fermi surface and a hole Fermi
surface that are nearly nested but do have
a finite curvature; moreover, it takes into account an electron reservoir
that mimics other bands. We determine the transport coefficients using a geometric
approach. We find that most transport coefficients display a leading dependence that
is linear in the energy gap. However, due to cancellation of the electron
and hole contributions, the Hall conductivity $\sigma_{xy}$ is essentially
unaffected by the onset of magnetism, even in the immediate vicinity of the quantum
critical point. The insensitivity of $\sigma_{xy}$ to magnetism is, therefore, a
robust feature of the model over a wide range of parameters across the magnetic
quantum phase transition. This conclusion is in agreement with the experiments
\cite{rosenbaum2,rosenbaum1}.

The remainder of the paper is organized as follows: In Section II, we discuss the
Fermi surface of Cr, and the Rice model used in this paper \cite{rice}. In Section
III, we discuss the changes in the electron spectrum upon magnetic ordering. In
Section IV, we describe a geometric picture \cite{ong} that affords a simple
description of the transport coefficients near the quantum critical point in
Cr$_{1-x}$V$_x$.  In Section V, we discuss numerical calculations based on the
actual band structure of Cr.  Some conclusions are offered in Section VI.
Calculational details are described in the Appendices.

\section{Model Fermi surface of chromium}\label{sec:modelFS}

\begin{figure}[t]
 \epsfxsize=8cm
 \epsfbox{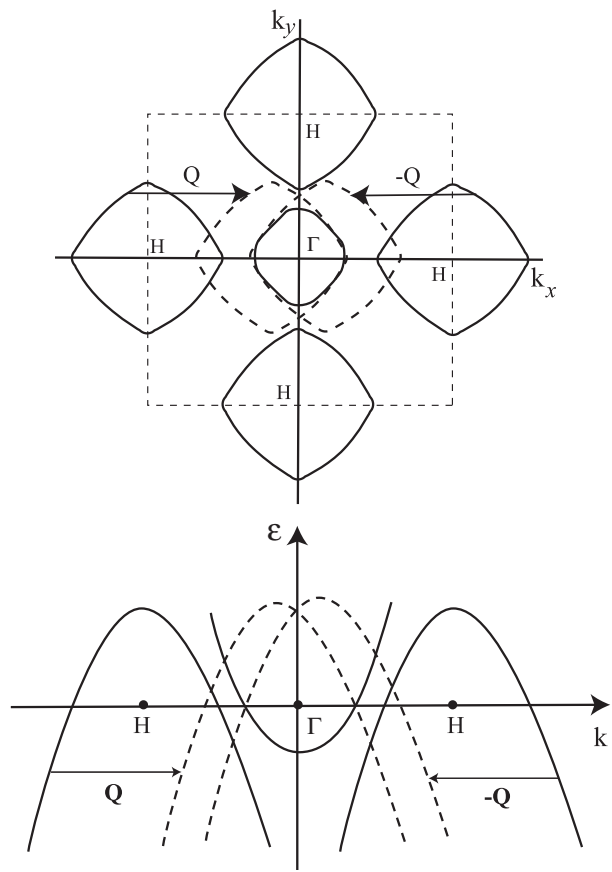}
\caption{Top: a 2D cross section of the relevant Fermi surface sheets in Cr.
The electron-like Fermi surface centered at the $\Gamma$-point and
the hole-like Fermi surface centered at the H-point have regions
that are almost nested. These regions lead to the SDW formation.
The dashed lines show the H-centered Fermi surface displaced
by the vectors $\pm {\bf Q}$. Bottom: schematic view of the electron
and the hole energy bands along the $\Gamma$-H line.
The dashed lines shows the hole band
displaced by $\pm {\bf Q}$.}
 \label{fig:chr_FS_sketch}
\end{figure}

The Fermi surface of chromium consists of several sheets and is well accounted for
by band structure calculations \cite{fawcett1,fawcett2}. It is believed that the
interaction between two particular sheets: the $\Gamma$-sheet, centered around the
$\Gamma$-point, and the H-sheet, centered around the H-point, is responsible for
the magnetism. These sheets are nearly nested; as a result, an SDW with nesting wave
vector ${\bf Q}$ emerges, as shown in Fig.~\ref{fig:chr_FS_sketch}. In chromium, the
SDW is of the linear type, which means \cite{elliott+wedgwood} that both $+{\bf Q}$
and $-{\bf Q}$ nesting is relevant.

The relative size of the Fermi surface (FS) sheets
can be changed by doping with vanadium. It is
believed that the dominant effect of the doping is to change the electron
concentration, which corresponds to the shift of the chemical potential $\mu$.
As $\mu$ shifts with increasing hole doping $x$, the mismatch between the
$\Gamma$-surface and the H-surface increases. The H-sheet grows and the
$\Gamma$-sheet diminishes in size. This leads to the change of wave vector with
concentration ${\bf Q}={\bf Q}(x)$.
At the critical doping $x_c$,
the mismatch becomes so large that the SDW formation is not energetically
favorable any more.

To allow for an analytic treatment, we will employ the Rice model of itinerant
antiferromagnetism \cite{rice} that involves an electron and a hole Fermi
surface of spherical shape and unequal size. The Fermi surface of chromium
is certainly far from being spherical \cite{our_chromium_PRL}.
However, it is not completely flat, and therefore the
nearly nested parts can be fairly accurately approximated by spherical segments.
It is these parts that are important in the development of the SDW. Also, as
long as the SDW gap is small, only these parts are destroyed by the
gap opening. This is why a model that considers a spherical FS will give
the correct behavior of physical quantities as $\Delta \to 0$. The highly
curved, non-nested
parts of the $\Gamma$-and H-sheets will introduce
corrections only when $\Delta$ becomes large.

We consider spherical Fermi surfaces of radii $k_{Fe}$ for the electron
surface and $k_{Fh}$
for the hole surface. Wherever possible
[for exceptions, see Secs.
\ref{sec:linear_conductivity_touching_spheres},
\ref{sec:hall-touching}],
we will linearize the spectrum:
\begin{eqnarray}
 \nonumber
 \epsilon_e(k) &=& v_e (|k| - k_{Fe})
 \\
 \label{eq:e12}
 \epsilon_h(k) &=& - v_h (|k| - k_{Fh})
\end{eqnarray}

\begin{figure}[t]
 \epsfxsize=8cm
 \epsfbox{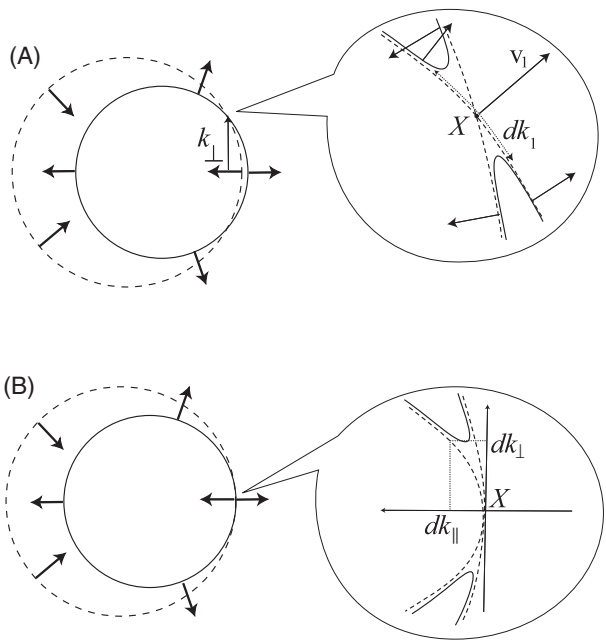}
\caption{The electron-like FS (solid line)
and the hole-like FS displaced by
$+{\bf Q}$ (dashed line).
The arrows show velocities at the Fermi surface.
They point outside the electron Fermi surface and inside the hole
Fermi surface. (A) the ``crossing'' case, (B) the ``touching'' case}
 \label{fig:FS_inner_positioning}
\end{figure}

We will present an analytic treatment of two cases
(see Fig.~\ref{fig:FS_inner_positioning}). In the first case,
upon translation by $\bf{Q}$, the electron Fermi surface
crosses the hole Fermi surface.
In the second case, the two Fermi surfaces touch each other
upon translation. The latter case is never realized
in the Rice model: the maximum in the susceptibility
is at a ${\bf Q}$ vector corresponding to crossing spheres \cite{rice}.

Note that in Fig.~\ref{fig:FS_inner_positioning},
the model hole surface is shown displaced only by the $+{\bf Q}$ vector,
while in the real situation of chromium (Fig.~\ref{fig:chr_FS_sketch})
both the $+{\bf Q}$ and the $-{\bf Q}$ displacements lead to nesting
and should be considered. This is done for simplicity, and also because
our analytic calculation is intended to give only the functional form
of the $\Delta$ dependence. Taking into account both
displacements changes only the value of the coefficient, not the leading
power of the dependence. In Appendix~\ref{appendix:linear_vs_helical},
we explain how the full result for the linear SDW is related to
that of the $+{\bf Q}$ displacement only.

There are other sheets of the Fermi surface of chromium that are not
shown in Fig.~\ref{fig:chr_FS_sketch}. They do not cause the SDW
formation, but are also modified by the onset of the SDW, and thus
their contribution to the transport coefficients changes as well.
The changes of
those Fermi surfaces
will correspond
to the crossing case of
Fig.~\ref{fig:FS_inner_positioning},
and our analysis applies to them as well,
as long as we are interested only in the
leading power in the dependence of transport
coefficients on the energy gap $\Delta$.

\section{spectrum in the ordered state}

Onset of a linear SDW generates a non-zero matrix ele\-ment $\Delta$ between the
single-electron states connected by the ordering wave vector $\pm{\bf Q}$. These
matrix elements are independent of spin~\cite{elliott+wedgwood} (as opposed to the
case of a helical SDW). As a result, a gap opens at all momenta ${\bf k}$ and ${\bf
k} \pm {\bf Q}$, related by the resonance condition $\epsilon({\bf k})=\epsilon({\bf
k} \pm {\bf Q})$, where $\epsilon({\bf k})$ is the single-particle spectrum in the
paramagnetic state. When the chemical potential falls inside the gap, some parts of
the original (paramagnetic state) Fermi surface disappear, and the FS sheets
reconnect as shown in Fig.~\ref{fig:FS_inner_positioning}.

In the following,
we neglect smaller gaps that appear whenever
two single-particle states meet a higher-order
resonance condition $\epsilon({\bf k})=\epsilon({\bf k}+n
{\bf Q})$ with $n>1$, since the magnitude of these higher-order gaps decreases
rapidly with $n$, as $\Delta_n \sim \Delta (\Delta/\epsilon_F)^{n-1}$.

In chromium,
the condition $\epsilon({\bf k})=\epsilon({\bf k} \pm {\bf Q})$
is  fulfilled only for electrons belonging to different
(hole-like and electron-like) bands. The same is true for the Rice model.
Thus the spectrum in the ordered state takes the form
\begin{eqnarray}
 \label{eq:E}
 \epsilon_{\pm}({\bf k}) &=& \frac{\epsilon_{1} + \epsilon_{2}}{2} \pm \sqrt{
 \left( \frac{\epsilon_{1} - \epsilon_{2}}{2} \right)^2 + \Delta^2 },
 \\
 \nonumber
 \epsilon_{1}({\bf k}) &=& \epsilon_e({\bf k}),
 \\
 \nonumber
 \epsilon_{2}({\bf k}) &=& \epsilon_h({\bf k} \pm {\bf Q}).
\end{eqnarray}
As discussed in the previous section, to find the functional dependence of physical
quantities on $\Delta$, we will use only one sign of ${\bf Q}$ in the formula for
$\epsilon_2$ (see Appendix~\ref{appendix:linear_vs_helical}). Thus we will
consider one electron band and one displaced hole band.

The original spherical Fermi surfaces of electrons and holes are given by the
equations
\begin{eqnarray}
 \nonumber
 \epsilon_{e,h}({\bf k}) &=& 0.
\end{eqnarray}
If the electron reservoir provided by other sheets of the Fermi surface is large,
the gap opening does not lead to a shift of the chemical potential. The new Fermi
surfaces are then given by
\begin{eqnarray}
 \label{eq:newFS}
 \epsilon_{\pm}({\bf k}) &=& 0,
\end{eqnarray}
As shown in Fig. \ref{fig:FS_inner_positioning}, when the SDW gap opens, the Fermi
surfaces reconnect. Instead of the two original spherical Fermi surfaces, one gets a
larger hole FS and a smaller lens-shaped electron FS. The latter corresponds to the
$(+)$ sign, while the former corresponds to the $(-)$ sign in Eq.~\ref{eq:newFS}. In
the touching case, there is no lens, thus only the $(-)$ sign is relevant. For
either sign, this equation is equivalent to
\begin{eqnarray}
 \label{eq:FS}
 \epsilon_1({\bf k}) \epsilon_2({\bf k})&=& \Delta^2.
\end{eqnarray}
This may be complicated in $k$-space, but is very simple in the space of
$(\epsilon_1,\epsilon_2)$.

The velocities in the modified bands are:
\begin{eqnarray}
 \label{eq:v}
 {\bf v}_{\pm} &=& \frac{1}{\hbar}
     \frac{\partial \epsilon_{\pm}}{\partial {\bf k}} =
 \\
 \nonumber
 &=&  \frac{{\bf v}_{1} + {\bf v}_{2}}{2}
 \pm  \frac{\epsilon_1 - \epsilon_2}{
           \sqrt{\left( \epsilon_1 - \epsilon_2 \right)^2 + 4\Delta^2}}
           \frac{{\bf v}_{1} - {\bf v}_{2}}{2}
\end{eqnarray}
In this formula, velocities ${\bf v}_{i}$ on the right hand side
are taken at the same momentum $\bf k$, so one or both of them
have to be evaluated far from the corresponding original electron
or hole Fermi surfaces.

Away from the points where the electron and the hole
Fermi surfaces cross or touch each other,
the new Fermi surface is close to one of them.
Quantitatively, this means
that the distance between the new and old surfaces
is much smaller then the distance between FS$_1$ and FS$_2$.
The corresponding condition reads (for definiteness,
we write it here for one of the branches):
 $$
 |\epsilon_{+} - \epsilon_1| << |\epsilon_{1} - \epsilon_2|,
 $$
which, using (\ref{eq:E}), can be re-expressed as
\begin{eqnarray}
 \nonumber
 |\epsilon_{1} - \epsilon_2| &>>& \Delta .
\end{eqnarray}
The crossover from large to small deviation
of the new Fermi surface from the old one takes place at
\begin{eqnarray}
    \label{eq:crossover}
    |\epsilon_{1} - \epsilon_2| &\approx& \Delta.
\end{eqnarray}

The above equations neglect the variation of the chemical potential $\mu$ upon onset
of ordering. The ordering wave vector $\bf{Q}$ also changes, and the dependences
$\Delta(x)$, $\mu(x)$ and $Q(x)$ on doping, $x$, need to be found
self-consistently. Within the framework of the Rice model, this was done
within a mean-field treatment in the
original paper \cite{rice}. Both $\mu(x)$ and $Q(x)$ turn out to depend on doping
linearly: $$ \mu(x) = \mu(x_c) + \mu^\prime \cdot (x - x_c), \,\,\, Q(x) =
Q(x_c) + Q^\prime \cdot (x - x_c), $$ while $\Delta(x)$ shows the standard
mean-field dependence $\Delta(x) \sim \sqrt{x_c - x}$. Hence, in the
immediate
vicinity of the transition, the leading dependence of the transport coefficients on
doping is defined implicitly by the dependence on $\Delta(x)$, and can be found
neglecting the variation of $\mu$ and $\bf{Q}$. We now turn to finding this
dependence.

\section{Transport coefficients}\label{sec:transport}

\subsection{Conductivity}

In the Boltzmann equation approach, the conductivity is given by the integral
\begin{eqnarray}
 \label{eq:sigmaIntegral}
 \sigma_{\alpha\beta} &=&
 \frac{e^2 \tau}{4 \pi^3 \hbar}
 \int v_{\alpha} v_{\beta} \frac{dS_F}{v_F}
\end{eqnarray}
where $v_{\alpha} = (1/\hbar) (\partial \epsilon_{\pm}/\partial k_{\alpha})$, $e$ is
the electron charge, $dS_F$ is the Fermi surface element in $k$-space, and
$\tau$ is the electron scattering time, hereafter assumed isotropic \cite{foot1}. In the SDW
phase, this expression differs from that in the paramagnetic state in two ways.
First, the integrand contains velocities defined by (\ref{eq:v}) and, second, the
integration is performed over the modified Fermi surface, given by (\ref{eq:FS}).

An analytic calculation is straightforward (see
Appendix~\ref{appendix:electric_conductivity}), but here we estimate the integral
(\ref{eq:sigmaIntegral}) using geometric considerations, which will allow us to
develop an intuitive picture.

\subsubsection{Crossing spheres}

First, we perform an estimate in the crossing case. When the gap opens, the original
Fermi surfaces disappear in belts of width $d k_{1,2}$ and radius $k_{\perp}$
determined by the positioning of the spheres (see
Fig.\ref{fig:FS_inner_positioning}(A). For clarity, only ${\bf v}_{1}$ and $d k_1$
are shown in the inset). The width of the disappearing belt can be evaluated using
(\ref{eq:crossover}). For an infinitesimal $\Delta$, it implies that the original
electron and hole Fermi surfaces are destroyed in belts of width
\begin{eqnarray}
    \label{eq:belts_size}
    dk_i &\approx& \frac{2|v_i|}{|{\bf v}_1 \times {\bf v}_2|}
\frac{\Delta}{\hbar} . \qquad (i = 1,2)
\end{eqnarray}
(see Appendix~\ref{appendix:electric_conductivity} for details).
This formula is valid only in the crossing case, where
${\bf v}_1 \times {\bf v}_2 \neq 0$.
Disappearance of the belts is accompanied by Fermi surface
reconnection, i.e. new portions of the FS are created. As seen from
Fig.\ref{fig:FS_inner_positioning}(A), the area
of the new FS is also proportional to $\Delta$.

Before the opening of the gap, the belts were giving a contribution of approximately
$(v_{\alpha,1}^2 dk_1 + v_{\alpha,2}^2 dk_2) 2 \pi k_{\perp}$ to
$\sigma_{\alpha\alpha}$. In this formula, the index $\alpha$ is the direction in
which the conductivity is measured. For the spherical Fermi surfaces considered
here, two distinct directions exist: one along the line connecting the centers of
the FS$_1$ and FS$_2$ spheres, and another one perpendicular to it. The
conductivities along these directions are denoted $\sigma_{||}$ and
$\sigma_{\perp}$, respectively. For both of them, the velocity projections are
zeroeth order in $\Delta$, thus the contribution of the belts is proportional to
their areas, i.e. first order in $\Delta$.  Fermi velocities on the reconnecting
parts of the FS are given by (\ref{eq:v}) and thus are also zeroeth order in
$\Delta$. Therefore the contribution of these parts is first order in $\Delta$ as
well.

Thus, when the belts disappear, and the reconnecting parts are created, the change
of $\sigma$ is
\begin{eqnarray}
 \label{eq:delta_sigma_estimate}
 \delta\sigma_{(||,\perp)} & \sim & \Delta.
\end{eqnarray}

\subsubsection{Touching spheres}\label{sec:linear_conductivity_touching_spheres}

Next, we estimate the change of conductivity in the touching case. Here,
(\ref{eq:FS}) implies that the old Fermi surfaces are destroyed in the circles
centered at the touching point (see Fig.\ref{fig:FS_inner_positioning}(B)).
In a coordinate system with axes along $\bf Q$ and perpendicular to it,
\begin{eqnarray}
 \nonumber
 \epsilon_1 & = & v_1 d k_{||} + \frac{d k_{\perp}^2}{2 m_1}
 \\
 \label{eq:touching_energy}
 \epsilon_2 & = & v_2 d k_{||} + \frac{d k_{\perp}^2}{2 m_2}
\end{eqnarray}
That gives for both FS$_{1,2}$
\begin{eqnarray}
 \nonumber
 d k_{||} \sim d k_{\perp}^2
\end{eqnarray}
Using now (\ref{eq:crossover}) and (\ref{eq:touching_energy}) to obtain the radius $d
k_{\perp}$ of the destroyed circle, we find
$d k_{\perp} \sim  \sqrt{\Delta}$.
The area
of the destroyed patches of the Fermi surfaces is $A \sim d k_{\perp}^2 \sim \Delta$.
The area of the reconnecting surface can be estimated as the area of a cylinder with a radius
of order $\sqrt\Delta$ and height $d k_{||} \sim d k_{\perp}^2 \sim \Delta$. Thus the
area of the reconnecting FS part is of order $\Delta^{3/2}$.

In contrast with the crossing case, the integrand of (\ref{eq:sigmaIntegral}) in the
touching case strongly depends on the direction $\alpha$. For $\sigma_{||}$, it is
simply equal to $(v_1^2 + v_2^2)$, i.e., zeroeth order in $\Delta$, so the
contribution of the destroyed part of the Fermi surface is $(v_1^2 + v_2^2)A \sim
\Delta$.  By virtue of the argument already used for the crossing case, the
integrand for the reconnecting part is also zeroeth order in $\Delta$. Since the
area of the reconnecting Fermi surface is higher order then the one of the
disappearing FS, its contribution can be neglected.  Therefore, the change of the
parallel conductivity is given by
\begin{eqnarray}
 \label{eq:delta_sigmapar_touching}
 \delta\sigma_{||} & \sim & \Delta.
\end{eqnarray}

For $\sigma_{\perp}$, the integrand of (\ref{eq:sigmaIntegral}) is much smaller,
because the projection of $v_F$ on the perpendicular direction vanishes as one
approaches the touching point. The characteristic value of the projection is $v_F
(dk/k_F) \sim v_F \sqrt\Delta$, thus the contribution of the disappearing part of the FS
to $\sigma_{\perp}$ is $(v_1^2 + v_2^2)\Delta A \sim \Delta^2$.  For the same
reason as in the calculation of $\sigma_{||}$, the value of the integrand on the
reconnecting parts of  the FS is of the same order $\sqrt\Delta$. Consequently, the
contribution of these parts is of higher order ($\sim\Delta^{5/2}$) and can be
neglected.  One obtains:
\begin{eqnarray}
 \label{eq:delta_sigmaperp_touching}
 \delta\sigma_{\perp} & \sim & \Delta^2.
\end{eqnarray}

Formulae
(\ref{eq:delta_sigma_estimate},\ref{eq:delta_sigmapar_touching},\ref{eq:delta_sigmaperp_touching})
can be compared with those obtained by Elliott and Wedgwood
\cite{elliott+wedgwood}. For the touching case, our results are identical to theirs,
while for the crossing case, Ref.~\onlinecite{elliott+wedgwood} gives $d\sigma_{||} \sim
\Delta$, $d\sigma_{\perp} \sim \ \Delta^2$. The discrepancy with
(\ref{eq:delta_sigma_estimate}) arises from the fact that
Ref.~\onlinecite{elliott+wedgwood} considers crossing of two spheres with equal
Fermi momenta and Fermi velocities, while in the Rice model, generically, the sizes
of the spheres differ. Coinciding Fermi momenta and Fermi velocities lead to the
cancellation of tems linear in $\Delta$, which cannot be captured by geometrical
reasoning. This point is further explained in Appendix~\ref{appendix:electric_conductivity}.

\subsection{Hall conductivity}

The Hall conductivity $\sigma^{(H)}_{ij}$ has the form $\sigma^{(H)}_{\alpha\beta} =
\sigma^{(H)}_{\alpha\beta\gamma} B_{\gamma}$, where, for small magnetic fields
~\cite{ziman},
\begin{eqnarray}
 \label{eq:hall_conductivity_classic}
 \sigma^{(H)}_{\alpha \beta} &=& \varepsilon_{\gamma \delta \sigma}
\frac{e^3 \tau^2}{4 \pi^3 \hbar c}
\int \frac{d S_F}{v_F}
        v_\alpha v_\sigma
        \frac{\partial^2 \epsilon}
             {\partial k_\beta \partial k_\delta} B_{\gamma}.
\end{eqnarray}
The change of this quantity upon onset of the SDW is found by calculating the
difference between the integrals above in the SDW state (with the modified Fermi
surface and Fermi velocities) and in the paramagnetic state. However, performing
such a calculation for the linearized spectrum (\ref{eq:e12}), one finds
$\delta\sigma^{(H)}_{\alpha\beta\gamma} = 0$. This is due to the fact that although
the curvature of the Fermi surface changes significantly upon reconnection of the
two Fermi surface sheets, the two connecting arcs give equal contributions of
opposite sign, which cancel each other exactly for a linear spectrum. Thus, one has
to expand $\epsilon_{1,2}$ up to quadratic terms in $\delta {\bf k}$ even in
the crossing case.

Instead of performing the calculation based on
(\ref{eq:hall_conductivity_classic}), one can find the leading power in the
dependence of $\delta\sigma^{(H)}_{\alpha\beta\gamma}$ on $\Delta$ in a more
intuitive way by using the approach of Ong \cite{ong}. In this method, the Fermi
surface is divided into slices perpendicular to the magnetic field $\bf B$. The
contribution of each slice to the Hall conductivity is given by the area $A_v$
inside the contour plot of ${\bf v}(k), \ k \in {\rm FS}$. For $B || z$,
\begin{eqnarray}
 \nonumber
 d\sigma^{(H)}_{xy} &=& \frac{e^2 \tau^2}{h} \frac{A_v}{\pi l_B^2},
\end{eqnarray}
where $ l_B = \sqrt{c\hbar/e B}$ is the magnetic radius, and the total Hall
conductivity is given by the integral
\begin{eqnarray}
 \label{eq:Ongs_sigmaH}
 \sigma^{(H)}_{xy}
 &=&
\frac{e^2 \tau^2}{h} \int \frac{d k_z}{2\pi} \frac{A_v(k_z)}{\pi l_B^2}.
\end{eqnarray}
Variation of $A_v(k_z)$ at each slice $k_z$ upon onset of the SDW can be found from
simple geometric considerations, which we describe below.

In the Rice model, the Fermi surface has cylindrical symmetry around the direction
of $\bf Q$. Accordingly, there are two distinct elements of the $\sigma^H$ tensor.
We denote them $\sigma^{(H)}_{||}$ and $\sigma^{(H)}_{\perp}$, where parallel and
perpendicular refers to the mutual orientation of $\bf Q$ and $\bf B$. In the
coordinate system where $\bf Q$ points along $\hat x$, the conductivity tensor in a
magnetic field has the form
\begin{eqnarray}
  \label{eq:sigma_tensor}
  \sigma &=& \left( \begin{array}{ccc}
    \sigma_{||} & \sigma^{(H)}_{\perp} B_z &  \sigma^{(H)}_{\perp} B_y
    \\
     & &
    \\
    -\sigma^{(H)}_{\perp} B_z & \sigma_{\perp} &  \sigma^{(H)}_{||} B_x
    \\
    & &
    \\
    -\sigma^{(H)}_{\perp} B_y & -\sigma^{(H)}_{||} B_x &  \sigma_{\perp}
    \end{array} \right).
\end{eqnarray}

Below, we find the functional forms of $\sigma^{(H)}_{||}(\Delta)$ and
$\sigma^{(H)}_{\perp}(\Delta)$.

\subsubsection{Crossing spheres}

\begin{figure}
 \epsfxsize=8cm
 \epsfbox{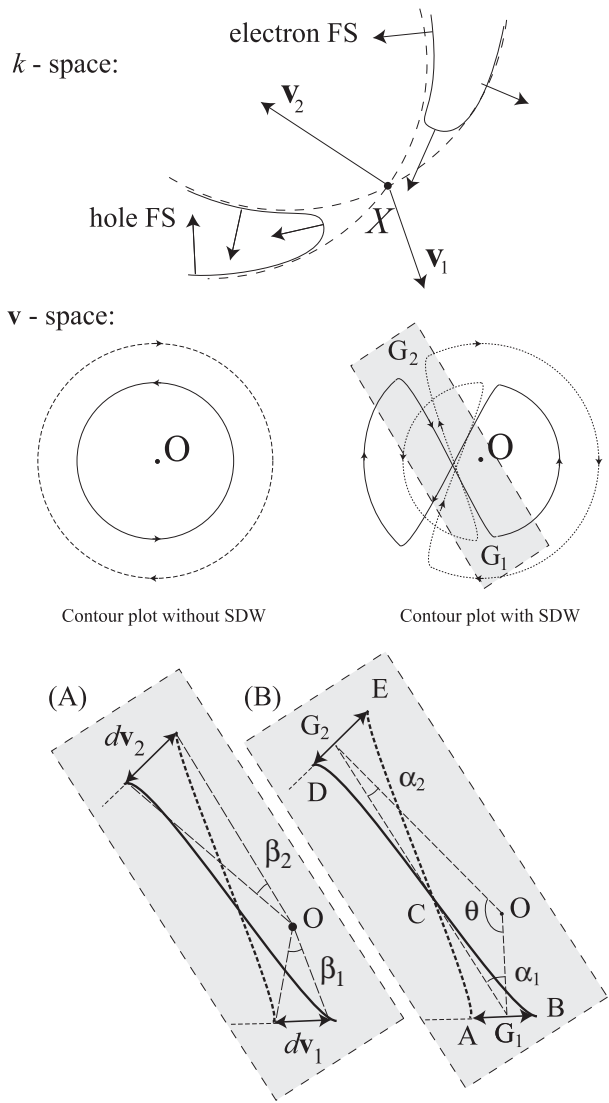}
\caption{Opening of the SDW gap modifies the Fermi surface ($k$-space) and the
contour plot of the velocity ($\bf v$-space). In the velocity contour plots (before
and after the SDW onset), the solid line corresponds to the electron and the dotted
line to the hole surface. Panels  (A) and (B) are blow-ups of the shaded
rectangle on the left side of the $\bf v$-space panel. They give the definitions of
the angles $\alpha_{1,2}$, $\beta_{1,2}$ and $\theta$. Point $O$ is the origin in
the $\bf v$-space. Points $G_{1,2}$ are the center points of the intervals $d{\bf
v}_{1,2}$. Points $A,B,D,E$ are defined as the points where deviations
occur from the paramagnetic case. They are defined by
the criteria (\ref{eq:crossover}).}
 \label{fig:v_contour_crossing}
\end{figure}

{\bf Calculation of $\sigma^{(H)}_{\perp}$ (${\bf B} \perp {\bf Q}$).} Let the $\hat
x$-axis point along ${\bf Q}$, and the $\hat z$-axis point along ${\bf B}$. The
spherical electron and hole Fermi surfaces intersect along a circle that lies in the
plane $k_x = {\rm const}$. Following Ong \cite{ong}, consider a cross-section of
$k$-space perpendicular to ${\bf B}$ ($k_z = {\rm const}$ cross-section). This
cross-section cuts the electron and the hole Fermi surfaces along the circles and,
depending on the value of $k_z$, these circles may or may not cross. When they do
cross, the contour plot of $\bf v$ is significantly modified due to opening of the
gap. A typical modification of the Fermi surface cross-section upon transition from
the paramagnetic to the SDW state is shown in Fig.~\ref{fig:v_contour_crossing}, as
well as the corresponding modification of the $\bf v$ contour plot. Note that, in
this plot, all $\bf k$ and $\bf v$ vectors are projections of the corresponding
three-dimensional vectors on the $k_z = {\rm const}$ plane. This means that the
diameters of the circles in the contour plot of $\bf v$ are not $v_1$ and $v_2$, but
instead are given by the absolute value of the projections. However, to keep
notation compact, we do not explicitly show this in the following discussion.

Near the crossing points $X$ in $k$-space (Fig.~\ref{fig:v_contour_crossing}, upper
panel), the Fermi surface cuts reconnect. In the $\bf v$-space, short segments
disappear in the $\bf v$ contour plots, originating from the electron and hole Fermi
surfaces, and those contours reconnect as shown in the middle right panel of
Fig.~\ref{fig:v_contour_crossing}. The connecting segments are close to straight
lines. Note that in formula (\ref{eq:Ongs_sigmaH}), the area inside the contour is
positive or negative, depending on the contour orientation. Examining the
reconnection pattern in Fig.~\ref{fig:v_contour_crossing}, one concludes that the
contribution $d\sigma^{(H)}$ under discussion can be estimated as the area
difference of the two curved elongated triangles - $ABC$ and $CDE$ in the blow ups
Fig.~\ref{fig:v_contour_crossing}(A,B).

The triangle sizes depend on $\Delta$. Let us first estimate the length of the the
triangle bases $d{\bf v}_{1,2}$, i.e. of the segments that disappear from the
original $\bf v$-space contour plots. In the $k$-space panel of
Fig.~\ref{fig:v_contour_crossing}, the length of the disappearing segments is given
by (\ref{eq:belts_size}). In the $\bf v$-space, we first find the angles
$\beta_{1,2}$ at which the gaps in FS$_{1,2}$ are seen from the centers of the
respective Fermi sphere (see Fig.~\ref{fig:v_contour_crossing}(A))
\begin{eqnarray}
 \nonumber
 \beta_i &=& \frac{dk_i}{k_{Fi}}.
\end{eqnarray}
The lengths of the disappearing segments in the contour plots are
\begin{eqnarray}
 \nonumber
 d{v}_i &=& v_{i} \beta_i = \frac{v_{i}}{k_{Fi}} dk_i,
 \\
 \nonumber
 d{v}_1 &=&  \frac{v_1^2}{|{\bf v}_1 \times {\bf v}_2|} \frac{2\Delta}{\hbar k_{F1}},
 \\
 \nonumber
 d{v}_2 &=& \frac{v_2^2}{|{\bf v}_1 \times {\bf v}_2|} \frac{2\Delta}{\hbar k_{F2}}.
\end{eqnarray}
Hence both triangles in Fig.~\ref{fig:v_contour_crossing}(A,B) have areas
proportional to $\Delta$. The dominant contribution to the Hall conductivity would
be $\delta\sigma^{(H)} \sim \Delta$, unless the triangles are of equal area, in
which case only the higher order terms in $\Delta$ survive the subtraction. Here we
show that this is not the case. Using the angles defined in
Fig.~\ref{fig:v_contour_crossing}(A,B), one finds the ratio of the triangle areas
\begin{eqnarray}
 \label{eq:triangle_area_ratio_1}
 \frac{A_1}{A_2} &=& \frac{dv_1 \cos\alpha_1}{dv_2 \cos\alpha_2} =
     \frac{\cos\alpha_1}{\cos\alpha_2} \frac{k_{F2} v_1^2}{k_{F1} v_2^2}.
\end{eqnarray}
Due to the fact that $\alpha_1$, $\alpha_2$, and $\theta$ are the angles in the
triangle $O G_1 G_2$ with the sides $|OG_1| = v_1$ and $|OG_2| = v_2$, everything
can be expressed through $\theta$
\begin{eqnarray}
 \label{eq:triangle_area_ratio_2}
 \frac{A_1}{A_2} &=&
     \frac{v_1 + v_2 \cos\theta}{v_2 + v_1 \cos\theta} \ \frac{k_{F2} v_1^2}{k_{F1} v_2^2}.
\end{eqnarray}
Thus, in the generic case,
the two areas are different, and our conclusion about
$\delta \sigma^{(H)} \sim \Delta$ holds.

The next step is to integrate over $k_z$. As mentioned earlier, some $k_z = {\rm
const}$ sections contain the crossing line of the electron and the hole Fermi
surfaces, and some do not. The former sections contribute $\delta \sigma^{(H)} \sim
\Delta$, while the latter sections contribute a higher power of $\Delta$, since
these contours deform only slightly. In the case of crossing spheres, the number of
contributing sections is proportional to $k_{\perp}$, and thus is independent of
$\Delta$. As a result, the change of the whole integral (\ref{eq:Ongs_sigmaH}) is of
the order of
\begin{eqnarray}
 \label{eq:delta_sigmaHperp_crossing}
 \delta\sigma^{(H)}_{\perp} \sim \Delta.
\end{eqnarray}

{\bf Calculation of $\sigma^{(H)}_{||}$ (${\bf B} || {\bf Q}$).} In this case, there
are no contours that deform significantly, but a set of contours totally disappears.
Each disappearing contour is of radius $k_{\perp}$, and contributes $\delta
\sigma^{(H)} \sim A \sim (k_{\perp}/k_F)^2$ to the Hall conductivity. The main
contribution to the integral over $k_x$ comes from the contours in the interval
$dk_{x} \sim \Delta$. Therefore, in the crossing case, and for $B||Q$, one finds
\begin{eqnarray}
 \label{eq:delta_sigmaHpar_crossing}
 \delta\sigma^{(H)}_{||} & \sim & \Delta.
\end{eqnarray}

\subsubsection{Touching spheres}
\label{sec:hall-touching}

{\bf Calculation of $\sigma^{(H)}_{\perp}$ (${\bf B} \perp {\bf Q}$).} The contour
plot transformation differs from the case of crossing spheres in three ways.

The first difference is that now the vanishing segments of the original contours
$d{\bf v}_1$ and $d{\bf v}_2$ are parallel to each other (in the notation of
equation (\ref{eq:triangle_area_ratio_1}), $\cos\alpha_1 = \cos\alpha_2 = 1$).

The second difference is that the area of each triangle is $A \sim \sqrt{\Delta}$.
This is because the length of the triangle bases $d v$ is proportional to the
length of the disappearing FS $d k_{\perp}$ and, as discussed in
Sec.~\ref{sec:linear_conductivity_touching_spheres} and shown in
Fig. \ref{fig:FS_inner_positioning},
$d k_{\perp} \sim \sqrt\Delta$. The areas of different sign still do not cancel
each other in the generic case, so each section now gives a contribution
$d\sigma^{(H)} \sim \sqrt\Delta$.

The third difference is that now the number of sections with significant
modification of the contours is also proportional to $d k_{z} = d k_{\perp} \sim
\sqrt\Delta$. Applying the same logic as in the crossing case, we conclude that
\begin{eqnarray}
 \label{eq:delta_sigmaHperp_touching}
 \delta\sigma^{(H)}_{\perp} & \sim & \Delta.
\end{eqnarray}

{\bf Calculation of $\sigma^{(H)}_{||}$ (${\bf B} || {\bf Q}$).} There are no
contours that deform significantly, but a set of contours disappear altogether. Each
vanishing contour had the radius $d k_{\perp} \sim \sqrt\Delta$ and contributed
$d\sigma^{(H)} \sim A \sim \Delta$. The main contribution to the integral over $k_x$
comes from the interval $d k_{x} \sim d k_{\perp}^2 \sim \Delta$ (again, see
discussion in Sec.~\ref{sec:linear_conductivity_touching_spheres}). As a result, in
the case of touching spheres,
\begin{eqnarray}
 \label{eq:delta_sigmaHpar_touching}
 \delta\sigma^{(H)}_{||} & \sim & \Delta^2.
\end{eqnarray}

\subsection{Hall coefficient}

The resistivity tensor is the inverse of (\ref{eq:sigma_tensor}). Calculated up to
the first order in $B$, it reads:
\begin{eqnarray}
  \nonumber
  \rho &=& \left( \begin{array}{ccc}
    \frac{1}{\sigma_{||}} &
          -\frac{\sigma^{(H)}_{\perp}}{\sigma_{||} \sigma_{\perp}} B_z &
                    -\frac{\sigma^{(H)}_{\perp}}{\sigma_{||} \sigma_{\perp}} B_y
    \\
     & &
    \\
    \frac{\sigma^{(H)}_{\perp}}{\sigma_{||} \sigma_{\perp}} B_z &
                 \frac{1}{\sigma_{\perp}} &
                       -\frac{\sigma^{(H)}_{||}}{\sigma^2_{\perp}} B_x
    \\
     & &
    \\
    \frac{\sigma^{(H)}_{\perp}}{\sigma_{||} \sigma_{\perp}} B_y &
                  \frac{\sigma^{(H)}_{||}}{\sigma^2_{\perp}} B_x &
                         \frac{1}{\sigma_{\perp}}
    \end{array} \right)
\end{eqnarray}
Consequently, two distinct Hall coefficients can be defined. For $\bf B$ along $\bf
Q$, and current perpendicular to both of them,
\begin{eqnarray}
 \nonumber
 R_{||} &=& \frac{\sigma^{(H)}_{||}}{\sigma^2_{\perp}},
\end{eqnarray}
and, for $\bf B$ perpendicular to $\bf Q$, and current perpendicular to $\bf B$,
\begin{eqnarray}
 \nonumber
 R_{\perp} &=& \frac{\sigma^{(H)}_{\perp}}{\sigma_{||} \sigma_{\perp}}.
\end{eqnarray}
Using equations (\ref{eq:delta_sigma_estimate}, \ref{eq:delta_sigmapar_touching},
\ref{eq:delta_sigmaperp_touching}, \ref{eq:delta_sigmaHperp_crossing},
\ref{eq:delta_sigmaHpar_crossing}, \ref{eq:delta_sigmaHperp_touching},
\ref{eq:delta_sigmaHpar_touching}) we get
\begin{eqnarray}
 \nonumber
 \delta R_{|| / \perp} \sim \Delta \qquad {\rm (crossing \ case)}
\end{eqnarray}
and
\begin{eqnarray}
 \nonumber
 \delta R_{||} &\sim& \Delta^2 \qquad {\rm (touching \ case)}
 \\
 \nonumber
 \delta R_{\perp} &\sim& \Delta
\end{eqnarray}

\section{Numerical Results}

For the case of vanadium doped chromium, our analytic results can be checked by
doing a numerical calculation based on the band theory Fermi surface of chromium.
Such a calculation was presented in our earlier paper \cite{our_chromium_PRL}, where
we emphasized the strong change in the Hall number with doping, as observed
experimentally \cite{rosenbaum1}.  However, we also implied that the actual
variation of the numerical results with doping near the critical point might be
different than that given by the analytic results based on a spherical Fermi
surface.  We did not, though, work close to the critical point because of
convergence issues.  This slow convergence can be traced to the problem of switching
from the antiferromagnetic $\epsilon_{\pm}$ basis to the paramagnetic
$\epsilon_{1,2}$ basis for the energy bands as the energy gap goes to zero, which as
we stated in the earlier paper, would lead to a discontinuity in the Hall
number for a perfectly flat Fermi surface.  Since the actual Fermi surface always
has some curvature, this problem can be compensated for by going to a more dense
breakdown of the Brillouin zone when doing the momentum integration using the linear
tetrahedron method.

We now describe how these calculations are performed.  Chromium has three energy
bands which cross the Fermi energy.  One of them defines a hole octahedron around
the $H$ point as well as hole ellipsoids around the $N$ points.  The second defines
an electron octahedron about the $\Gamma$ point.  The third defines electron balls
centered around the $\Gamma-H$ lines.  (In this paper, we ignore the third band when
computing the transport integrals, thus the electronic structure becomes of the same
two band form, as described in the previous sections.)  The paramagnetic band
structure is constructed from the local density approximation using a linear muffin
tin orbital method with muffin tin overlap corrections, and spin-orbit treated in a
semi-relativistic fashion.  After convergence, eigenvalues were generated at 506 $k$
points in the irreducible (1/48th) wedge of the Brillouin zone. These were then fit
by a 910 function Fourier series (spline fit), which allows eigenvalues to be
generated for an arbitrary $k$.  An SDW gap is then introduced as in the previous
sections.  Because of the convergence issues mentioned above, the calculation was
carried out to even higher precision than in our previous work.  That is, the
transport integrals are performed using the linear tetrahedron method with the
momentum sums involving over 805 million tetrahedra in the paramagnetic zone.

\begin{figure}[t]
 \epsfxsize=3.4in
 \epsfbox{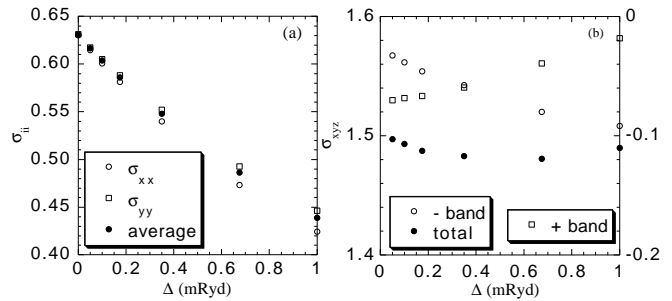}
\caption{(a) Calculation of $\sigma_{||} (\sigma_{xx})$ and $\sigma_{\perp}
(\sigma_{yy})$ for chromium as a function of the energy gap, $\Delta$ (milli-Rydberg
units).  The chemical potential was set to correspond to four percent hole doping,
with a $\bf Q$ vector of $2\pi/a(0.909,0,0)$. Empty symbols represent the result of
a calculation in the single domain state. Filled symbols are the domain averaged
result. (b) Calculation of $\sigma_{xyz}$ for the domain averaged case. Although the
small variation in each band (+,-) is linear in $\Delta$, the signs are opposite,
leading to a nearly flat behavior of the total $\sigma_{xyz}$ with respect to
$\Delta$.}
 \label{fig:numeric}
\end{figure}

In Fig.~\ref{fig:numeric}(a), we show results for the longitudinal ($\sigma_{||}$) and
transverse ($\sigma_{\perp}$) conductivities (assuming a single $Q$ domain).
The variation is indeed linear in $\Delta$ for both cases, in
agreement with the analytic results (crossing case).

The $\Delta$-dependence of $\sigma^{(H)}_{xyz}$ is shown in
Fig.~\ref{fig:numeric}(b). The value of $\sigma^{(H)}_{xyz}$ hardly changes as the SDW gap
$\Delta$ (equivalently, the magnetic order parameter) sets in.  This is a
consequence of the flatness of the octahedral surfaces along with the
electron-hole compensation of the $\Gamma$ and $H$ sheets in the regions
where they nest (the large absolute
magnitude of $\sigma^{(H)}_{xyz}$ is mostly coming from the $N$ ellipsoids, which
are essentially unaffected by the magnetism).

We can, though, test the analytic dependence derived earlier by looking at
the individual contributions of the + and - bands, which we also plot in
Fig.~\ref{fig:numeric}(b).  A clear linear $\Delta$ dependence is
seen.  Interestingly, the two linear terms have opposite sign with almost
equal magnitude, which is a reflection of the electron-hole compensation in
this basis.

We have also tested our argument further by numerically determining
$\sigma^{(H)}_{xyz}$ for the Rice model with unequal-sized spheres. Our numerical
results indeed verify the analytic results presented earlier in this paper.
Moreover, we find that even in the paramagnetic case,
$\sigma_{xyz}$ is small due
to near electron-hole compensation, and also has a weak dependence on
$\Delta$ for the same reason.  As a consequence, $\sigma_{xyz}$ is primarily determined
by the ``reservoir", which stays essentially unchanged as the magnetic order sets
in. Indeed, as we mentioned above, the dominant contribution to $\sigma^{(H)}_{xyz}$
plotted in Fig.~\ref{fig:numeric}(b) comes from the $N$ ellipsoids.

As a consequence, we find the deviation of the Hall number to be linear in $\Delta$
as well, in agreement with the analytic results.  Moreover, the near constancy of
$\sigma^{H)}_{xyz}$ implies that the Hall number should scale with the square of the
longitudinal conductivity, a point discussed in our previous paper
\cite{our_chromium_PRL}.

We have also checked the dependence of $\sigma_{ii}$ on the variation {\bf q} of the
ordering wave vector {\bf Q}, and found it to be of order {\bf q}$^2$, thus a higher
order effect.  We note that {\bf Q} itself was chosen so as to yield the maximum in
the susceptibility (that is, optimal nesting), the latter calculated using the same
linear tetrahedron method employed in the transport calculations.
In addition, the occupation number change with $\Delta$ is small, meaning the
chemical potential variation with $\Delta$ is also a higher order effect.
Therefore, the linear variation of $\sigma_{ii}$ and $R_H$ with $\Delta$ we find
both numerically and analytically appears to be a robust result.

\section{Conclusions}

In conclusion, we have analytically estimated the longitudinal and the Hall
conductivities for a spin density wave metal. Our conclusions are two-fold. First,
we find that, under generic circumstances, these transport coefficients depend
linearly on the SDW energy gap in the small gap limit.  These results have been
confirmed by numerical calculations using the actual Fermi surface of chromium.

Second, we showed that the behavior of the Hall conductivity $\sigma^{(H)}_{xyz}$ is
special in the case of V-doped Cr in that the change $\delta\sigma^{(H)}_{xyz}$ upon
magnetic ordering is small. This is due to the fact that the nested regions of
the hole and electron surfaces which are removed by magnetic ordering give
contributions which are approximately equal in magnitude but opposite in sign,
as revealed by detailed numerical calculations based on the actual Fermi surface.

The effect of this electron-hole compensation is particularly obvious within the Rice
model, where the critical mismatch of the electron and the hole surfaces is a small
number of order $\Delta(0)/\epsilon_F$, $\Delta(0)$ being the value of the energy
gap for perfect nesting, and $\epsilon_F$ being the Fermi energy. Hence, the total
Hall conductivity of the two bands, and its change upon the gap opening, is reduced
by a factor of order $\Delta(0)/\epsilon_F$ compared with the contribution of each
band.  In both the case of the Rice model, and the true Fermi surface of chromium,
the value of $\sigma_{xyz}$ itself is largely set
by the electron reservoir, i.e. by electrons that are unaffected by the magnetism.
As a consequence,
$\sigma^{(H)}_{xyz}$ is nearly constant over a wide parameter range across the
quantum critical point.

We now make a brief comparison with the experiments \cite{rosenbaum1,rosenbaum2}.
The latest pressure results in V-doped Cr \cite{rosenbaum2} have provided access
to the behavior in the immediate vicinity of the quantum critical point.
The Hall coefficient was found to have a square root dependence on pressure: $\delta
R_H \sim \sqrt{p_c - p}$.  The N\'eel temperature also has the same dependence
and, since within mean field theory, $\Delta$ is proportional to $T_N$, we conclude
that $\Delta \sim \sqrt{p_c - p}$ as well.
Thus $R_H$ indeed depends linearly on $\Delta$, as we found theoretically.
Experiment also finds that the Hall conductivity $\sigma^{(H)}_{xyz}$ is unchanged
across the quantum critical point, and that $\rho_{xy}$ scales with $\rho_{xx}^2$,
as we also expect.

However, the longitudinal resistivity was found to vary as $\delta\rho \sim (p_c -
p)^{2/3}$.\cite{rosenbaum2} This is different from our theoretical prediction
$\delta\rho \sim \Delta \sim \sqrt{p_c - p}$ and seems to be at variance with the
claim that $\rho_{xy} \sim \rho_{xx}^2$.
We believe this is connected with the difficulty of extracting critical exponents in
the presence of sample inhomogeneity, though we cannot exclude the possibility of
new physics beyond the mean-field picture.

The overall success of the nested SDW picture for understanding
the $T=0$ data in vanadium-doped chromium shows that the Hall
effect is indeed a good measure of Fermi surface evolution across
a quantum critical point. Our conclusion is not only important for
simple metals such as chromium, but also serves as a benchmark
for the understanding of the Hall effect in more complex systems
such as heavy fermion metals \cite{PASCHEN} and high ${\rm T_c}$
cuprates \cite{Balakirev}.

Finally, we should stress that our theory only applies at zero
temperature, and does not address the strong temperature dependence
of the Hall conductivity observed experimentally \cite{rosenbaum1}.
We leave these interesting challenges for future theoretical studies.

\acknowledgments

We would like to thank Tom Rosenbaum, Gabe Aeppli, Anke Husmann, and Minhyea Lee for
communicating their unpublished results to us. This work was supported by the U. S.
Dept. of Energy, Office of Science, under Contract No. W-31-109-ENG-38 (MRN, YB, RR)
and by NSF Grant No.\ DMR-0090071, Robert A. Welch Foundation, and TCSAM (QS).

\appendix

\section{}\label{appendix:electric_conductivity}

To simplify notation, we introduce
\begin{eqnarray}
 \nonumber
 \xi &=& (\epsilon_1 - \epsilon_2)/2,
 \\
 \nonumber
 E_0 &=& \sqrt{ \xi^2 + \Delta^2}.
\end{eqnarray}
Then
\begin{eqnarray}
 \nonumber
 \epsilon_{\pm} &=& \frac{\epsilon_1 + \epsilon_2}{2} \pm E_0,
\end{eqnarray}
and, for the velocities in the modified bands, we find:
\begin{eqnarray}
 \label{eq:appendix_v}
 {\bf v} &=& \frac{1}{2}({\bf v}_{1} + {\bf v}_{2}) \pm
           \frac{\xi}{2 E_0} ({\bf v}_{1} - {\bf v}_{2}),
\end{eqnarray}
where $ {\bf v}_{i} = v_i \frac{\bf k}{|k|}$. Since the integral in
(\ref{eq:sigmaIntegral}) is taken over the modified Fermi surface, we first have to
evaluate expression (\ref{eq:appendix_v}) at the new FS. We start with the electron
FS (equation (\ref{eq:v}) with the ``$+$'' sign) for which (\ref{eq:appendix_v}) is
equivalent to
\begin{eqnarray}
 \label{eq:v_on_FS}
 {\bf v} &=& u^2 {\bf v}_{1} + v^2 {\bf v}_{2},
 \\
 \nonumber
 u^2 & \equiv & \frac{1}{2} \left( 1 + \frac{\xi}{E_0} \right),
 \\
 \nonumber
 v^2 & \equiv & \frac{1}{2} \left( 1 - \frac{\xi}{E_0} \right),
\end{eqnarray}
where the factors $u^2, v^2$ are mathematically identical to the BCS coherence
factors. At the Fermi surface, one can express all quantities through $\epsilon_1$
using (\ref{eq:FS}). Keeping in mind that for the electron FS $\epsilon_1 < 0$, we
calculate
\begin{eqnarray}
 \nonumber
 \xi &=& \frac{\epsilon_1}{2} \left(1 - \frac{\Delta^2}{\epsilon_1^2} \right),
 \\
 \nonumber
 E_0 &=& \sqrt{\xi^2 + \Delta^2} = \sqrt{\left(\frac{\epsilon_1}{2} \right)^2
         \left(1 + \frac{\Delta^2}{\epsilon_1^2} \right)^2 }
 \\
 \label{eq:appendixA_E0}
         &=&
         \left| \frac{\epsilon_1}{2} \right|
         \left(1 + \frac{\Delta^2}{\epsilon_1^2} \right) =
         - \frac{\epsilon_1}{2}
         \left(1 + \frac{\Delta^2}{\epsilon_1^2} \right)
\end{eqnarray}
Now one can express the $u^2, v^2$ factors at the FS:
\begin{eqnarray}
 \nonumber
 u^2 &=& \frac{\Delta^2}{\epsilon_1^2 + \Delta^2},
 \\
 \label{eq:u2v2_on_FS}
 v^2 &=& \frac{\epsilon_1^2}{\epsilon_1^2 + \Delta^2}.
\end{eqnarray}
Thus, for the electron FS,
\begin{eqnarray}
 \label{eq:v_on_FS_2}
 {\bf v}_{F} &=&  \frac{\Delta^2}{\epsilon_1^2 + \Delta^2} {\bf v}_{1} +
 \frac{\epsilon_1^2}{\epsilon_1^2 + \Delta^2} {\bf v}_{2},
 \\
 \nonumber
 {v}_{F}^2 &=& \frac{ \Delta^4 {v}_{1}^2 + 2 \Delta^2 \epsilon_1^2 ({\bf
 v}_{1} \cdot {\bf v}_{2}) + \epsilon_1^4 {v}_{2}^2}{(\epsilon_1^2 +
 \Delta^2)^2}.
\end{eqnarray}
Note, however, that in this equation ${\bf v}_{1,2}$ must be evaluated on the FS, so
if the velocities vary substantially, the usage of (\ref{eq:v_on_FS_2}) is not
straightforward.

For the hole FS there is a sign change in (\ref{eq:v_on_FS}). Also now $\epsilon_1 >
0$ and there is an additional sign change in the formula (\ref{eq:appendixA_E0}) for
$E_0$. As a result, the formula for $\bf v$ comes out to be the same. Thus
Eq.~(\ref{eq:v_on_FS_2}) is valid for both parts of the FS.

The expression for the conductivity  (\ref{eq:sigmaIntegral}) is written in cartesian
coordinates $(k_x, k_y, k_z)$. However, it is much easier to evaulate it in the
$(\epsilon_1,\epsilon_2,\phi)$ coordinates, $\phi$ being the angle of rotation around
the $\bf Q$ axis, because of the simplicity of Eq. (\ref{eq:FS}) in these
coordinates. To do that, we need to express the element $d S_F$ of the Fermi surface
area in the $(\epsilon_1,\epsilon_2,\phi)$ coordinates.

Fig.~\ref{fig:FS_inner_positioning}(A) shows a section of the electron and hole
Fermi spheres in a plane cutting through their centers, i.e. a $\phi = {\rm const}$
section. All $\phi = {\rm const}$ sections are identical in our model. In the plane,
we introduce cartesian coordinates $(k_{||}, k_{\perp})$ with $k_{||}$ being along
$Q$. Using a linear approximation for the energies $\epsilon_1, \epsilon_2$ near the
point $X$, where FS$_1$ and FS$_2$ cross, one has
\begin{eqnarray}
 \nonumber
 \epsilon_1 &=& {\bf v}_1 \cdot (\hbar d{\bf k}),
 \\
 \label{eq:linear_e12}
 \epsilon_2 &=& {\bf v}_2 \cdot (\hbar d{\bf k}),
\end{eqnarray}
with ${\bf v}_{1,2}$ taken at point $X$. It is convenient to introduce two
reciprocal vectors ${\bf u}_{1,2}$, such that
\begin{eqnarray}
 \nonumber
 {\bf v}_i \cdot {\bf u}_{j} &=& \delta_{ij}.
\end{eqnarray}
Explicitly,
\begin{eqnarray}
 \nonumber
 {\bf u}_{1}
 &=&
 \frac{ {\bf v}_{2}
\times
[ {\bf v}_{1} \times {\bf v}_{2}] }{
            ( {\bf v}_{1} \times {\bf v}_{2} )^2 },
 \\
 \label{eq:u12}
 {\bf u}_{2}
 &=&
 \frac{ {\bf v}_{1}
 \times
 [ {\bf v}_{2} \times {\bf v}_{1}] }{
            ( {\bf v}_{1} \times {\bf v}_{2} )^2 }
\end{eqnarray}
In terms of ${\bf u}_{1,2}$, one has
\begin{eqnarray}
 \label{eq:dpvec}
 d{\bf k} &=& {\bf u}_{1} d\epsilon_1 + {\bf u}_{2}d\epsilon_2
  = \big(  {\bf u}_{1} - \frac{\Delta^2}{\epsilon_1^2} {\bf u}_{2} \big)
          \frac{d\epsilon_1}{\hbar},
\end{eqnarray}
where the second expression is valid only at the Fermi surface. Squaring equation
(\ref{eq:dpvec}), we obtain
\begin{eqnarray}
 \nonumber
 dk^2 &=&
 \frac{ {\bf v}_2^2 + 2 (\Delta^2/\epsilon_1^2)
({\bf v}_1 \cdot {\bf v}_2) +
    (\Delta^4/\epsilon_1^4)
{\bf v}_1^2 }{( {\bf v}_{1}
 \times
 {\bf v}_{2} )^2 } \
    \left(\frac{d\epsilon_1}{\hbar}\right)^2.
 \\
 \label{eq:dp2_crossing}
 & &
\end{eqnarray}
Using (\ref{eq:v_on_FS_2}) and (\ref{eq:dp2_crossing}) together, we find
\begin{eqnarray}
 \nonumber
 \frac{d k}{|v_F|} &=& \frac{\epsilon_1^2 + \Delta^2}{\epsilon_1^2}
        \frac{(d\epsilon_1/\hbar)}{|{\bf v}_1 \times {\bf v}_2 |},
 \\
 \nonumber
 \frac{d S_F}{|v_F|} &=&  \frac{\epsilon_1^2 + \Delta^2}{\epsilon_1^2}
        \frac{(d\epsilon_1/\hbar)}{|{\bf v}_1 \times {\bf v}_2 |} k_{\perp} d\phi,
\end{eqnarray}
where $k_{\perp}$ is the radius of the circle over which the two original FS cross.
If the integrand has no $\phi$ dependence, one can write
\begin{eqnarray}
  \label{eq:integral_over_FS}
  \int \frac{d S_F}{|v_F|}
 &=&
\frac{2\pi k_{\perp}}{|{\bf v}_1 \times {\bf v}_2 |}
           \int_{-\infty}^{+\infty} \frac{d\epsilon_1}{\hbar} \frac{\epsilon_1^2 + \Delta^2}{\epsilon_1^2}.
\end{eqnarray}
Here integration over the $(-\infty,-0)$ interval accounts for the electron FS
contribution, and integration over $(+0, +\infty)$ accounts for the hole FS
contribution.

We now continue with the calculation of the conductivity tensor
$\sigma_{\alpha\beta}$. With our choice of coordinates, it is diagonal, and its two
unequal components are $\sigma_{||}$ and $\sigma_{\perp}$. To use integral
(\ref{eq:sigmaIntegral}), note that
\begin{eqnarray}
 \nonumber
 v_{||}^2 &=& (u^2 v_{1||} + v^2 v_{2||})^2,
 \\
 \label{eq:v_components}
 \left(
 \begin{array}{c} v_{\perp 1}^2 \\ \\ v_{\perp 2}^2 \end{array}
 \right)
 &=& (u^2 v_{1\perp} + v^2 v_{2\perp})^2
 \left(
 \begin{array}{c} \sin^2\phi \\ \\ \cos^2\phi \end{array}
 \right).
\end{eqnarray}
Start with $\sigma_{||}$. Using (\ref{eq:u2v2_on_FS}) in (\ref{eq:v_components}) we
can now write down the expression for the integrand of (\ref{eq:sigmaIntegral}). Up
to a constant factor $e^2 \tau/(4\pi^3\hbar)$ one gets:
\begin{eqnarray}
 \nonumber
 \sigma_{||} & \sim &
           \frac{2\pi k_{\perp}}{|{\bf v}_1 \times {\bf v}_2 |} \times
 \\
 \nonumber
 & \times &
           \int \frac{d\epsilon_1}{\hbar} \frac{\epsilon_1^2 + \Delta^2}{\epsilon_1^2}
           \left\{
         u^4 v_{1||}^2 + 2 u^2 v^2 v_{1||}v_{2||} +
         v^4 v_{2||}^2
           \right\} =
 \\
 \label{eq:sigma_par}
 &=&
 \frac{2\pi k_{\perp}}{|{\bf v}_1 \times {\bf v}_2 |} \times
 \\
 \nonumber
 & \times &
           \int \frac{d\epsilon_1}{\hbar} \frac{1}{\epsilon_1^2 + \Delta^2}
    \left\{
         \frac{\Delta^4}{\epsilon_1^2} v_{1||}^2 +
         2 \Delta^2 v_{1||}v_{2||} +
         \epsilon_1^2 v_{2||}^2
    \right\}.
\end{eqnarray}
Notice that this integral does not converge near $\epsilon_1 = 0$. This only means
that our approximation (\ref{eq:linear_e12}) does not allow us to calculate the total value
of $\sigma$, since it is accurate only in the vicinity of the crossing point $X$.
However we can calculate the change of the conductivity, $\delta\sigma$. To do that,
we need to subtract the result corresponding to the paramagnetic case from
(\ref{eq:sigma_par}): the sum of the integrals over the original FS$_{1,2}$ (i.e.
along the dashed lines on Fig.~\ref{fig:FS_inner_positioning}(A)).

Up to the same constant factor $e^2 \tau/(4\pi^3\hbar)$ that we chose not to show
explicitly in (\ref{eq:sigma_par}), the contribution of FS$_2$ in the paramagnetic
state is
\begin{eqnarray}
 \nonumber
 \sigma^{(2)}_{||}& \sim & \frac{2\pi k_{\perp}}{|{\bf v}_1 \times {\bf v}_2|}
           \int \frac{d\epsilon_1}{\hbar} v_{2||}^2,
\end{eqnarray}
where $d\epsilon_1 v_2/(\hbar |{\bf v}_1 \times {\bf v}_2|) = d k_2$ is exactly the
length element of FS$_2$. By the same token, the FS$_1$ paramagnetic
contribution is
\begin{eqnarray}
 \nonumber
 \sigma^{(1)}_{||}& \sim & \frac{2\pi k_{\perp}}{|{\bf v}_1 \times {\bf v}_2|}
           \int \frac{d\epsilon_2}{\hbar} v_{1||}^2 =
     \frac{2\pi k_{\perp}}{|{\bf v}_1 \times {\bf v}_2|}
           \int \frac{d\epsilon_1}{\hbar} \frac{\Delta^2}{\epsilon_1^2}  v_{1||}^2.
\end{eqnarray}
We get
\begin{eqnarray}
 \nonumber
 \delta\sigma_{||} & \sim & \frac{2\pi k_{\perp}}{|{\bf v}_1 \times {\bf v}_2|}
           \int \frac{d\epsilon_1}{\hbar} \frac{1}{\epsilon_1^2 + \Delta^2}
    \left\{
         \frac{\Delta^4}{\epsilon_1^2} v_{1||}^2 + \right.
  \\
  \nonumber
  & & \left.
         + 2 \Delta^2 v_{1||}v_{2||} +
         \epsilon_1^2 v_{2||}^2
    \right\} - \frac{\Delta^2}{\epsilon_1^2}  v_{1||}^2 - v_{2||}^2 =
  \\
  \nonumber
  &=& \frac{2\pi k_{\perp}( 2 v_{1||}v_{2||} - v_{1||}^2 - v_{2||}^2)}
     {|{\bf v}_1 \times {\bf v}_2|}
     \int_{-\infty}^{+\infty}
     \frac{d\epsilon_1}{\hbar}\frac{\Delta^2}{\epsilon_1^2 + \Delta^2} =
  \\
  \label{eq:delta_sigma_par}
  &=& - \frac{2\pi k_{\perp}( v_{1||}- v_{2||})^2}
     {|{\bf v}_1 \times {\bf v}_2|} \frac{\Delta}{\hbar} \pi
\end{eqnarray}
Note the factor with the difference of Fermi velocity projections in this
formula. For the special case of identical Fermi surfaces, it may vanish for some directions, and
then the change of longitudinal conductivity will be
quadratic in $\Delta$. This
indeed happens in the case considered in Ref.~\onlinecite{elliott+wedgwood} and was
discussed in the body of the paper at the end of
Sec.~\ref{sec:linear_conductivity_touching_spheres}. In the Rice model
with different Fermi surfaces, such a cancellation does not happen.

For the other components of the conductivity, we merely need to use the
corresponding velocity projections in (\ref{eq:sigmaIntegral}). An additional
complication here is the $\phi$-dependence in (\ref{eq:v_components}), but it only
gives a factor of $1/2$:
\begin{eqnarray}
 \nonumber
 \delta\sigma_{\perp} & \sim & \frac{k_{\perp}}{|{\bf v}_1 \times {\bf v}_2|}
           \int d\phi \sin^2\phi
           \int \frac{d\epsilon_1}{\hbar}  \left\{ \ldots \right\} =
  \\
  \label{eq:delta_sigma_perp}
  &=& - \frac{\pi k_{\perp}( v_{1\perp}- v_{2\perp})^2}
     {|{\bf v}_1 \times {\bf v}_2|} \frac{\Delta}{\hbar} \pi.
\end{eqnarray}
The conclusion for the crossing case is that
\begin{eqnarray}
 \label{eq:appendix_delta_sigma_estimate}
 \delta\sigma_{(||,\perp)} & \sim & \Delta
\end{eqnarray}

\section{}\label{appendix:linear_vs_helical}

In the case of a linear SDW, each sheet of the FS interacts with the other one
shifted both by $+\bf{Q}$ and $-\bf{Q}$. For an incommensurate $\bf{Q}$, this leads
to an infinite secular matrix which has to be solved.  On the other hand, for Fermi
surface restricted integrals, we can formulate a good approximation which avoids
this difficulty. For the Rice model in the incommensurate case, for instance, it
never happens that the copies of the hole Fermi surface shifted by $+\bf{Q}$ and
$-\bf{Q}$ are both close to the same point of the electron Fermi surface. This
property considerably simplifies the calculation, because it allows us to consider
the hole FS shifted only by $+\bf{Q}$. Indeed, the electron Fermi surface can be
divided into three parts: (I) the one which disappears due to interaction with the
hole Fermi surface shifted by $+\bf{Q}$, (II) the one which disappears due to the
interaction with the other hole surface shifted by $-\bf{Q}$, and (III) the one
which survives. Any kinetic coefficient will be given by part III. At the same time,
when only one of the shifted surfaces is considered, it is given by a sum of parts
II and III for $+\bf{Q}$, and parts I and III for $-\bf{Q}$. Since the paramagnetic
value is a sum of the contributions from I, II, and III, it follows that the linear
SDW leads to coefficients which are equal to the sum of the values for the $+\bf{Q}$
and the $-\bf{Q}$ cases minus the paramagnetic value. Therefore, the change of the
kinetic coefficients is twice the result obtained in the calculation with the hole
surface shifted by only $+\bf{Q}$, since the $+\bf{Q}$ and $-\bf{Q}$ contributions
are equal by inversion symmetry.

A similar argument can be applied to the flat Fermi surface model of Shibatani {\it
et al} \cite{shibatani}, as well as to the actual Fermi surface of Cr.  In Section
V, the above relation was considered when doing the numerical calculations as well.


\begin{thebibliography}{2}
\bibitem{foot1}
Here we have assumed that the change of $\tau$ due to the magnetic
ordering can be ignored. This is expected to be valid for
intermediate values of the impurity scattering potential. It is
also experimentally found \cite{rosenbaum2} to be the case for
the pressure-tuned transition in Cr$_{1-x}$V$_x$
\end{thebibliography}
\end{document}